\documentstyle[12pt,world_sci]{article}
\catcode`@=11 \@addtoreset{equation}{section} \catcode`@=12

\newfont{\blackb}{msbm10 scaled 1200}
\newfont{\blacks}{msbm10 scaled 1000}  
\newfont{\blackl}{msbm10 scaled 1440}  
\def\BC{\mbox{\blackb C}} 
\def\BR{\mbox{\blackb R}} 
 
\def\BP{\mbox{\blackb P}} 

\def\CP{\mbox{\BC\BP}}

\begin{document}
\rightline{\tt hep-th/9703170}
\title{\Large Punctures in W-string theory}
\author{Suresh Govindarajan\thanks{Talk presented at the
Workshop on Frontiers in Field Theory, Quantum Gravity and String
Theory, Puri, December 12-21, 1996.} \\[-3pt]
{\it Department of Physics}\\[-3pt]
{\it Indian Institute of Technology}\\[-3pt]
{\it Madras 600 036 INDIA}\\[-3pt]
{\tt E-mail: suresh@imsc.ernet.in}}
\maketitle
\begin{abstract}
Using the differential equation approach to W-algebras, we discuss the 
inclusion of punctures in W-string theory. 
The key result is the existence of different kinds of punctures in
W-strings. We obtain the moduli associated with these punctures and
present evidence in existing W-string theories for these punctures.
The $W_3$ case is worked out in detail. It is conjectured that the
$(1,3)$ minimal model coupled to two dimensional gravity corresponds to
topological $W_3$-gravity. 
\end{abstract}
\renewcommand{\large}{\normalsize}
\renewcommand{\Large}{\normalsize}
\addtocounter{footnote}{1}
\section{Introduction}

Drinfeld and Sokolov\cite{ds} have constructed generalisations of the KdV
equation and its related integrable hierarchy. The basic input in
their construction is a semi-simple Lie group G and a principal
embedding of SL$(2)$ in G. Given this data, they constructed a linear
(pseudo-) differential operator from which the generalised KdV
hierarchy can be derived. The space of these (pseudo-) differential
operators has been shown to have a bi-Hamiltonian structure by Gelfand
and Dikii. It was soon 
realised that classical W-algebras arise precisely from one of these
Hamiltonian structures. (See  ref. \citenum{reviews} for more details as well
as references.) W-string theories can be defined to be those string
theories whose underlying chiral algebras are W-algebras.

In a seemingly unrelated development, Hitchin constructed generalisations of 
the Teichm\"uller spaces of Riemann surfaces\cite{hitchina,hitchinb}. 
The basic input was 
similar to that of Drinfeld and Sokolov -- a semi-simple Lie algebra G and a
principal embedding of SL$(2)$ in G. {\em Is there a relation between these two
constructions?}. This relation was established in ref. \citenum{proposal}
where it was shown that the generalised Teichm\"uller spaces of
Hitchin are indeed the Teichm\"uller spaces for the W-strings related
to the W-algebras obtained from the Drinfeld-Sokolov
construction\footnote{In the Polyakov path integral formulation of the
bosonic string, for a given genus $g$, the path integral over the two
dimensional metric reduces to a finite dimensional integral over the
moduli space of Riemann surfaces of genus $g$. This moduli space is
obtained from the Teichm\"uller space by the action of a discrete
group called the mapping class group.}. This connection also provides
a Polyakov path integral description for the ``W-gravity'' sector of a
W-string theory.

This however is not the end of the story. The geometric structures
(such as complex and projective structures) associated with the
generalised Teichm\"uller spaces of Hitchin are not known unlike the
case of the usual Teichm\"uller space which  arises as the space of
complex structures on a Riemann surface. 
In ref. \citenum{unif} a higher dimensional generalisation of the
Riemann surface was constructed to provide geometric insight into the
generalised Teichm\"uller spaces of Hitchin.  These manifolds will be
referred to as {\it W-manifolds}. As yet, not much is known about
these manifolds. One important property that is however known is that
W-diffeomorphisms linearise on W-manifolds. Thus this space plays a role
similar to the one played by superspace in the context of
supersymmetry. There have been other attempts at constructing geometric
structures related to W-gravity. Some recent attempts are refs.
\citenum{falqui} and \citenum{zucchini}.

The results in ref. \citenum{unif} are abstract since Hitchin's result require
the genus $g>1$.The case of $g=0,1$ being excluded by a stability
condition required on the Higgs bundle used in Hitchin's
construction\cite{hitchinb}. To obtain simpler examples, one needs to 
include the sphere as well as the torus.  For these cases, such a stability
condition can be satisfied if one includes punctures in the
construction. The results which I will describe in this workshop
addresses how to describe punctures on W-manifolds and discuss the
relationship of these results to W-string theory.  Some of the work
being reported here has been done in collaboration with
T. Jayaraman. A detailed report will appear elsewhere\cite{punctures}.
The main result being reported here is the existence of different kinds
of punctures in W-string theory.

The plan of the talk is as follows. The differential equation and the
related integrable hierarchy play a central role in the construction
of W-manifolds. This is described in sec. 2. In sec. 3, we
show how to incorporate punctures in this construction and show the
existence of many punctures in W-string theory. For explicitness, we
shall use the example of the $W_3$-string. We shall see that the 
$W_3$-string has two types of punctures. In sec. 4, we provide evidence for
these different punctures in existing W-string
theories. Finally, we conclude in sec. 5 with some observations.

\section{Constructing W-manifolds}

The starting point for the construction of W-manifolds is to consider
the linear homogeneous differential equation (possibly obtained from the
construction of Drinfeld and Sokolov) associated with the W-algebra of interest. 
When one considers the group G=SL$(n)$, one obtains the following operator
\begin{equation}
L=d^n+u_2(z)~d^{n-2}+\cdots+ u_n(z) \quad,
\end{equation}
where $d\equiv {d\over{dz}}$ and the corresponding differential equation
\begin{equation}
L~f=0\quad. \label{ediff}
\end{equation}
Here $z$ is a coordinate on some chart on a Riemann surface $\Sigma$ of
genus $g$. The $u_i$ are assumed to be such that the differential equation is
Fuchsian (i.e. the Frobenius method works) and has no regular singular 
points.

Locally (i.e. on a coordinate chart),
the linear differential equation (\ref{ediff}) has $n$ linearly
independent solutions. Let $\{f_i\}$ $i=1,\ldots,n$ be a basis. The
Wronskian is defined by
\begin{equation}
W=\left|\matrix{f_1 &\cdots & f_n \cr
                 f_1^{'}& & f_n^{'} \cr
	        \vdots &\ddots  & \vdots \cr
                 f_1^{(n-1)}&\cdots& f_n^{(n-1)}} \right|\quad.
\end{equation}
The non-vanishing of $W$ implies that the basis is linearly independent.
The vanishing of the coefficient of $d^{(n-1)}$ in the differential
equation (\ref{ediff}) implies that $W$ is a constant independent of
$z$. The basis can be normalised such that $W=1$. The basis $\{f_i\}$
can be interpreted as homogeneous coordinates on $\CP^{n-1}$. Thus
locally, solving the differential equation provides us with a map from
the coordinate chart into $\CP^{n-1}$. (More on the global properties of
this map later)

Globally, things are a little more intricate and interesting. First, for
the constant condition on the Wronskian to make global sense, we need
that $W$ transform like a scalar. This in turn implies that $f$
transforms as a $(1-n)/2$ differential. Since we are ``solving'' the
differential equation on a Riemann surface, there exist non-trivial
loops. One can analytically continue the basis $\{f_i\}$ along a
non-trivial loop $\gamma$. Then, the $f_i$ mix amongst each other
as follows
\begin{equation}
{f_i}^{'} = {M_i}^j(\gamma)~f_j\quad, \label{emonod}
\end{equation}
where $M$ is called the monodromy matrix associated with the loop
$\gamma$. For the differential equation we are considering, clearly,
the matrix $M(\gamma)\in$G=SL$(n,\BC)$. The analyticity of $f$ implies that the
monodromy matrix can only depend on the homotopy class $\pi_1(\Sigma)$
 of the loop $\gamma$. By means of analytic continuation of the basis,
we can associate to each element of the homotopy class $\pi_1(\Sigma)$,
a matrix $M\in G$. This forms the {\it monodromy group} of the
linear differential equation (the group action is the one induced by the
composition of loops). The freedom of choice of the basis $\{f_i\}$ implies
that this group is defined upto overall conjugation in SL$(n,\BC)$.

The map from $\Sigma$ to $\CP^{n-1}$ mentioned earlier globally is a
multi-valued (polymorphic) map. This multi-valuedness precisely encodes 
the monodromy data of the differential equation. A closed loop on
$\Sigma$ will get mapped to a open segment whose endpoints are related
by the monodromy matrix (Note that SL$(n,\BC)$ has a natural action on
$\CP^{n-1}$.). $\Sigma$ can be represented as the quotient of the upper half 
plane (which is the universal cover of $\Sigma$)
 by a Fuchsian sub-group of SL$(2,\BR)$ as follows from the
uniformisation theorem. Then, the multi-valued map becomes a
{\it single-valued map} from the upper half plane to $\CP^{n-1}$ with the image 
being  co-dimension $(n-2)$ surface.

This image can be extended to obtain a $(n-1)$ dimensional image by augmenting 
the differential equation by means of $(n-2)$ times of the 
generalised KdV hierarchy.  The times are defined by the equations
\begin{equation}
{d\over {dt_p}} L = [(L^{p/n})_+,L]\quad,
  \label{ekdv}
\end{equation}
where $p=2,\ldots,(n-1)$ and by $(L^{p/n})_+$, we mean the differential
operator part of the pseudo-differential operator $(L^{p/n})$. Thus the
$u_i$'s which entered the differential equation are now functions of
the KdV times. The importance of the times  of the generalised 
KdV hierarchy is that these furnish
{\it isomonodromic deformations} of the differential equation, i.e., they 
do not alter the monodromy group of the differential equation. 
	This extended image on being quotiented by the monodromy group
provides a higher dimensional generalisation of Riemann surfaces. The
moduli space of complex structures is related to the generalised
Teichm\"uller spaces of Hitchin. We refer to these objects as {\it
W-manifolds}. The non-linear W-diffeomorphisms turn out be a sub-group
of linear diffeomorphisms on W-manifolds. See refs. \citenum{gomis,unif} 
for more details.
	This construction is abstract and  as mentioned earlier, not
much is known about these manifolds.  Thus it is of interest
to obtain things in a more explicit fashion. One way to do this is to
include the cases of genus 0 and 1 into the construction of W-manifolds.
However, in order to obtain stable configurations (in the sense of Higgs
bundles in Hitchin's construction) one needs to include punctures. This
is addressed in the next section.

\section{Describing Punctures}

The differential equation approach provides a rather simple way of
including punctures into the construction of W-manifolds. Punctures are
defined as {\it regular singular points with specified monodromy conjugacy
class}. In the case of Riemann surfaces, the monodromy associated with a
puncture is parabolic, i.e., it is conjugate to $\pmatrix{1&1\cr0&1}$ 
and is an element of SL$(2,\BR)$.
This imposes the following condition on $u_2(z)$\cite{hempel}
\begin{equation}
u_2(z) = {1\over{4z^2}} + \cdots \quad,
\end{equation}
where $z=0$ is a puncture. The monodromy matrix around the puncture is
described by two real parameters and this is the moduli associated with
the puncture.

We shall now consider of the third order differential equation  which is
relevant for the $W_3$ case. Again, we shall define a puncture to be
a regular singular point and specified monodromy conjugacy class. What
are the allowed monodromies? Using the second order case as a guide, 
we shall consider unipotent matrices -- there is more than one of them.
We find that there are two distinct possibilities:
\begin{equation}
P=\pmatrix{1&1&0\cr0&1&1\cr0&0&1}\quad {\rm and} \quad
Q=\pmatrix{1&0&1\cr0&1&0\cr0&0&1}\quad,
\end{equation}
with the requirement that the monodromy matrix is in SL$(3,\BR)$.
The moduli associated with these punctures are obtained by the following
procedure.  A typical monodromy matrix of type $P$ is of the
form $APA^{-1}$ for some $A\in$SL$(3,\BR)/{\cal I}$, where ${\cal I}$ is the 
subgroup of $SL(3,\BR)$ matrices which commute with $P$. One can check that the
dimension of ${\cal I}$ is $2$ and hence the number of (real) parameters 
associated with a
$P$ type puncture is $8-2=6$. Similarly, the number of (real) parameters
associated with a $Q$-type puncture is $4$.
 
Thus, the differential equation approach to W-geometry suggests that
there exists two types of punctures with (complex) moduli three and two
respectively. The moduli associated with a genus $g$ surface with
$N_P$ P-type punctures and $N_Q$ Q-type punctures is given by 
\begin{equation}
(8g-8+3N_P+2N_Q)\quad. \label{emoduli}
\end{equation}
It is of interest to look for rigid objects, i.e., those without any
moduli. These occur at genus zero for $N_P=2$, $N_Q=1$ (this configuration 
will be represented by the symbol $\langle PPQ \rangle$) and for 
$N_P=0$, $N_Q=4$ (this configuration 
will be represented by the symbol $\langle QQQQ \rangle$). The choice of
these symbols is to emphasise the similarity to a topological
model to be discussed in the next section.

For the case of $W_n$-strings, there are $(n-1)$ types of punctures with
(complex) moduli ranging from $(n^2-n)/2$ to $(n-1)$.

\section{Evidence for the existence of different kinds of punctures}

The results of the previous section suggest the existence of two kinds
of punctures in $W_3$ string theory. Is this a reasonable suggestion? 
Different kinds of punctures have already occurred in superstring theory
-- the NS and R types. Thus this is not a radical suggestion.
We shall provide direct evidence for this from existing $W_3$-strings!
In addition, we shall also show how a certain existing topological model
is the perfect candidate for topological $W_3$-gravity.

\subsection{$W_3$ string theory}

The section makes extensive use of results of P. West and his collaborators.
Please see ref. \citenum{west,pope} for more details and related references. 
We shall now consider the $W_3$-string associated with pure
$W_3$-gravity. There is considerable evidence that this theory is
closely related to the Ising model coupled to 2d gravity\cite{ising}. 
The fields in this theory are:
\begin{itemize}
\item {\it Gravity sector:} the spin-2 ghost system $(b,c)$ and the spin-3
ghost system $(d,e)$.
\item {\it Liouville sector:} a scalar $\phi$
\item {\it Matter sector:} Scalars $x^\mu$ (for pure $W_3$ gravity, there is one scalar
field)
\end{itemize}
The states are labeled by the primaries of 
the Ising model -- $1,\sigma,\epsilon$ and are
\begin{eqnarray}
V(1,0)&=& c\partial e~e~e^{i\beta(1;0)\phi}~V^x(1)\quad,\\
V(\sigma,0)&=& c\partial e~e~e^{i\beta(\sigma;0)\phi}~V^x({{15}\over{16}})\quad,\\
V(\epsilon,0)&=& (c~e-{i\over{522}} \partial e~e)~
e^{i\beta(\epsilon;0)\phi}~V^x({1\over2})
\end{eqnarray}
where by $V^x(h)$, we mean a vertex operator in the matter sector with
dimension $h$; $\beta(1;n)=(8-8n)iQ/7$,
$\beta(\sigma ;n)=(7-4n)iQ/7$, $\beta(\epsilon;n)=(4-8n)iQ/7$ and
$Q^2=49/8$. The first two states given above occur at ``standard''
ghost number three while the last one (labeled by $\epsilon$) occurs at 
ghost number two. This unusual occurance was first observed by S. K.
Rama\cite{rama}.

In the operator formulation of string theory, the number of moduli  are
given by the number of antighost insertions required to obtain a
non-vanishing amplitude. Here there are two types of anti-ghosts  $b$
and $d$ given by the spin two and three systems respectively. Thus, the
ghost number of a state provides a direct count of moduli associated
with a state. We thus obtain that states labeled by $1$ and $\sigma$
have three moduli per insertion (and thus correspond to $P$-type punctures)
and states labeled by $\epsilon$ have two moduli per insertion (and thus
correspond to $Q$-type punctures). 

This simple minded counting can be seen to be correct by studying
scattering amplitudes in the theory\cite{fw}. One can look for states at
non-standard ghost numbers and this leads to an infinite set of states
labeled by
\begin{equation}
V(1,n)\quad;\quad V(\sigma,n)\quad;\quad   V(\epsilon,n) \label{enew}
\end{equation}
{\it Are these new states?} As we shall see now, these cannot be
considered as new states. Define the following operators
\begin{eqnarray}
\widehat{S} &\equiv& \oint \{d -{{5i}\over{3\sqrt{58}}} \partial b +
\cdots \} e^{i\beta^s\phi}\quad,\\
\widehat{P}&\equiv& [a\cdot x +\phi, Q_{BRST}]\quad,
\end{eqnarray}
where $\beta^s=-2iQ/7$ and $[\widehat{S},Q_{BRST}]=0$. One can show that
$V(\sigma,n)= (\widehat{S}^2\widehat{P})^n V(\sigma,0)$ with similar
relations holding for the other operators too. The form of $\widehat{S}$
shows that the insertion of a $\widehat{S}$ operator is equivalent
to carrying out a $\int d$ insertion. By similar arguments, one can show
that the operators given in eqn. (\ref{enew}) are not new states. 

Thus one can choose to work
with $n=0$ states and put in as many insertions of $\widehat{S}$ and
$\widehat{P}$ as are necessary.  Consider a scattering amplitude
involving $N_1$, $N_\sigma$ and $N_\epsilon$ states of type $1$,
$\sigma$ and $\epsilon$ respectively. The ghost number and Liouville
background charge conditions give us the numbers $N_S$ and $N_P$,
 of $\widehat{S}$ and $\widehat{P}$ insertions. The number of extra moduli
(in comparison with the bosonic case)
required are given by the expression $(N_S-N_P)$ which using the
expressions given by West work out to $(2N_1+2N_\sigma+N_\epsilon-5)$.
Adding the bosonic moduli, we get that the total moduli counting
works out to be $(3N_1+3N_\sigma+2N_\epsilon-8)$.
 Discounting the $-8$ in
the expression (this is due to zero modes of the ghosts on the sphere),
we see that there are three moduli per puncture for $1$ and $\sigma$
operators and two moduli per puncture for the $\epsilon$
operator\footnote{P. West interprets $(N_S-N_P+N_\epsilon)$ as the number
of extra moduli. Our interpretation removes this distinction between the
$\epsilon$ operator and the other operators.}. Comparing the moduli
count with that in eqn. (\ref{emoduli}), we obtain that the $1$ and
$\sigma$ correspond to punctures of type $P$ and $\epsilon$ corresponds
to punctures of type $Q$.

\subsection{A conjecture}

It is of interest to construct a topological model for $W_3$-gravity.
Such a model would probe the moduli spaces associated with
$W_3$-gravity. Rather than directly construct this model, we shall show
that an existing topological theory\footnote{W. Lerche has pointed out that 
there is evidence to suggest that certain W-minimal
models couples to W-gravity give rise to topological W-matter coupled to
topological W-gravity\cite{lerche}. We thank him for the information.} 
-- the $(1,3)$ minimal model
coupled to two dimensional gravity. This model is closely related to the
Gaussian point of the two-matrix model. This has been studied in some detail 
by Dijkgraaf and Witten\cite{wd}.  

We shall provide some relevant details of this model. The reader is
referred to the reference cited earlier for more details.  The {\it small
phase space} of this model consists of two operators labeled $P$ and
$Q$, in the notation of Witten and Dijkgraaf. At the topological point,
the only non-vanishing correlators at genus 0 are $\langle PPQ \rangle$
and $\langle QQQQ \rangle$. In topological theories, ghost charge is
related to the number of moduli. Further, at the topological point,
the only non-vanishing objects are those which saturate ghost charge
which implies that they have no moduli associated with them. This is
exactly what we had seen earlier with regard to the punctures of type
$P$ and $Q$. 

In continuum Liouville theory, one has vertex operator representations
of these operators. They are given by\cite{multi}
\begin{eqnarray}
P&=& e^{-3\phi_L/2\sqrt3}\ e^{iX/2\sqrt3} \quad,\\
Q&=& e^{-2\phi_L/2\sqrt3}\quad,
\end{eqnarray}
with the Liouville background charge $Q_L=-8/2\sqrt3$. The Liouville
charge conservation condition is
$$ \sum_i p_L^i = Q_L $$
where $p_L^i$ are the Liouville charges of the $i-th$ vertex operator in
a correlation function. Clearly,  $\langle PPQ \rangle$
and $\langle QQQQ \rangle$ saturate this condition and the Liouville charges of
the $P$ and $Q$ operators are in the ratio $3:2$ as required from our
moduli argument.

This shows that this model has all the right properties to be a theory
of topological $W_3$-gravity. It has two operators in the small phase
space which implies that it has two types of punctures. Further, the
moduli counting also works out right. Finally, this model exhibits
multicritical behaviour and the first critical point corresponds to
the Ising model coupled to gravity which we saw is related to $W_3$
gravity\cite{wd,multi}. This might suggest the following conjecture: the
$(1,n)$ model coupled to two dimensional gravity should represent
topological $W_n$ gravity. However, from $n=4$ onwards, the rigid
objects in this model are different from the one obtained from the 
differential equation approach pursued in this paper! Thus for $n>3$, 
the $(1,n)$ model coupled to two dimensional gravity does not correspond to
topological $W_n$-gravity. 

The topological theory probes the moduli spaces of $W_3$ gravity. Is
there something known about these spaces? Hitchin's result has to be
extended to include punctures. This extension has been carried out for
the case of $P$-type punctures\cite{parabolic}. However, so far the 
Higgs bundle corresponding to $Q$-type 
punctures have not been constructed. 

\section{Conclusion and Outlook}

In this talk, we have seen the existence of many different kinds of
punctures in W-string theory. The genus zero case can be studied in
greater detail and new insight may possibly be obtained. For example,
the new moduli which appear in W-strings may be understood better. In
the bosonic string, the four punctured sphere has one modulus which is
described by means of a cross-ratio. Thus all four point functions
depend on the positions of the punctures in terms of this cross-ratio. 
In the $W_3$-case, the first non-trivial modulus appears at genus zero 
for $N_P=0$ and $N_Q=6$. This must be described by means of a
generalised cross-ratio. Goncharov has constructed a generalised
cross-ratio\cite{goncharov} in proving certain conjectures of Zagier 
connecting
trilogarithms to zeta functions. This cross-ratio seems to have the right
properties (such as transformations under $SL(3)$),  and hence is the
candidate for the new W-modulus. This  also suggests an 
interesting interplay between Algebraic Number Theory and correlation
functions  in $W_3$ conformal field theory.

Some open problems with regard to the W-manifolds include the
construction of Higgs bundles which encompass all types of punctures,
more global information about these manifolds like the metric etc.
Since the W-manifolds seem to be higher dimensional generalisation of
Riemann surfaces which are the worldsheets for strings, it is natural
to think of these as the worldsheets for p-branes. If this were true, then
these manifolds will provide a topological expansion analogous
to the genus expansion for strings. 

One of the big problems with working with W-symmetries, is their
non-linearity. The linearisation provided by the W-manifolds should
provide some simplification in realising W-symmetries and hence
might lead to the construction of many more systems realising this symmetry. 

\noindent {\bf Acknowledgements:} We would like to thank the organisers
of the Puri workshop for the invitation and excellent hospitality.
We would like to thank T. Jayaraman and W. Nahm for several
conversations and P. West for useful correspondence explaining his work.

\end{document}